\documentclass[a4paper]{PoS}
\usepackage{mathtools}
\newcommand{\tr}{\text{tr}\,}
\usepackage{graphicx}
\usepackage{subcaption}

\title{Contribution title}

\title{Renormalization constants of the lattice energy momentum tensor
  using the gradient flow \footnote{Preprint number: CERN-PH-TH-2015-283}}

\ShortTitle{Lattice E.M.T. renormalization constants using the gradient flow}

\author{\speaker{Francesco Capponi}\\
Centre for Mathematical Sciences,
 Plymouth University,
 Drake Circus, Plymouth, PL4 8AA, U.K\\
  E-mail: \email{francesco.capponi@plymouth.ac.uk}}

\author{Luigi Del Debbio\\
 School of Physics and Astronomy, University of Edinburgh, Edinburgh, EH9 3JZ, Scotland, UK  \\
  E-mail: \email{luigi.del.debbio@ed.ac.uk}}

\author{Agostino Patella \\
Centre for Mathematical Sciences,
 Plymouth University,
 Drake Circus, Plymouth, PL4 8AA, U.K\\
  PH-TH, CERN, CH-1211 Geneva 23, Switzerland\\
  E-mail: \email{Agostino.Patella@cern.ch}}

\author{Antonio Rago\\
 Centre for Mathematical Sciences,
 Plymouth University,
 Drake Circus, Plymouth, PL4 8AA, U.K\\
  E-mail: \email{antonio.rago@plymouth.ac.uk}}

\abstract{We employ a new strategy for a non perturbative determination of the renormalized energy momentum tensor. The strategy is based on the definition of suitable lattice Ward identities probed by observables computed along the gradient flow.
The new set of identities exhibits many interesting qualities, arising from the UV finiteness of flowed composite operators.
In this paper we show how this method can be used to non perturbatively renormalize the energy momentum tensor for a SU(3) Yang-Mills theory, and report our numerical results.
}

\FullConference{The 33rd International Symposium on Lattice Field Theory\\
		14 -18 July 2015\\
		Kobe International Conference Center, Kobe, Japan}

\begin{document}

\section{Introduction}

The renormalization of the energy momentum tensor plays a crucial role in many different aspects of quantum field theory.
It is needed for correctly determining the energy, pressure and entropy density of a system at thermal equilibrium, or the shear viscosity of a fluid medium.
It can also be used as an order parameter for probing fundamental properties of quantum field theories, like conformal invariance.
The energy momentum tensor is related to the Noether current of translational symmetry:
in quantum field theory, the invariance of a physical system under this symmetry results in a collection of Ward identities that can be used to define a renormalized energy momentum tensor \cite{CCJ,CJ,Fujikawa}.

When we use a lattice regulator, translational invariance is broken and its restoration is only guaranteed in the continuum limit.
For this reason, at finite lattice spacing,  the correctly renormalized energy momentum tensor is obtained as a mixing of all operators with mass dimension less or equal than four, allowed by the lattice symmetries \cite{CCMP1,CCMP2}

\begin{equation}
(\hat{T}_{\mu \nu})_R=\sum_i Z_i \left \{\hat{T}_{\mu \nu}^{(i)}-\langle \hat{T}_{\mu \nu}^{(i)}\rangle \right\} \quad.
\end{equation}

In this proceedings we study a pure gauge theory with color group $SU(3)$.
The assumption that translational invariance is recovered in the continuum implies that it is possible to tune the coefficients $Z_i$
in such a way that the translation Ward identities are satisfied up to terms that vanish when the lattice spacing goes to zero. 
Unfortunately, local Ward identities are plagued by additional divergences (contact terms) which make the extraction of the coefficients $Z_i$ numerically more challenging. 

This problem can be solved with the aid of Yang Mills gradient flow, or Wilson Flow on the lattice \cite{Lwflow,Lwflow1,LWwflow2,Lwflow3}.
Using the smoothing properties of the flow, we can define a new set of
(local) translation Ward identities that do not contain any kind of contact term.
We show how these identities can be used to numerically measure the renormalization constants $Z_i$ and report preliminary results.
For a precise derivation of the equations used in this proceedings , we refer the reader to \cite{DDPR}.

Two alternative strategies have been recently proposed for the renormalization of the energy momentum tensor.
The first one makes use of the small flow time expansion of composite operators, built along the flow, in order to extract the expectation values of the renormalized energy momentum tensor \cite{Sz1,Sz2,Sz3,Sz4}. 
The second one employs shifted boundary conditions in order to define a set of suitable Ward identities that can be used for the numerical evaluation of the coefficients $Z_i$ \cite{GP,GP1}.

\section{Gradient Flow}

The Gradient Flow is defined by the following set of equations

\begin{equation}\label{flow_eq}
\begin{cases}
\partial_t B_{t,\mu}(x)=D_{t,\nu}G_{t,\nu \mu}(x) \quad,\\
 B_{0,\mu}(x)=A_\mu(x)\quad . 
\end{cases}
\end{equation}
Here $D_{t,\nu}$ and $G_{t,\nu \mu}$ represent the covariant derivative and strength tensor built with the field $B_{t,\mu}$ while $t$ is an additional parameter, with dimension of length squared, known in the literature as flow time.
At finite lattice spacing, a discretized form of equation (\ref{flow_eq}) can be defined as well \cite{Lwflow}. Starting from a gauge configuration $U_\mu(x)$ at flow time zero, using the discretized flow equation, we define a flowed gauge field $V_{t,\mu}(x)$ at any given time $t$.
The field $V_{t,\mu}(x)$ can be used for building any kind of gauge invariant composite operator whose expectation value we are interested in.
The remarkable property of these flowed operators is that they have a finite continuum limit for any non zero $t$, i.e. they do not require renormalization.

\section{TWI at positive flow time}

According to \cite{DDPR}, it is possible to define a set of Ward identities for translational invariance (TWI) using probe observables at positive flow time

\begin{equation}\label{flow_TWI}
\partial_\mu\langle [\hat{T}_{\mu \rho}]_R(x)\phi_t(y)\rangle=-Z_{\delta}\langle \hat{\delta}_{x,\rho}\phi_t(y)\rangle+O(a^2)\quad.
\end{equation}
Because of the UV finiteness of the probes, we do not have any kind of contact term at positive flow time.
Beside the renormalization constants encoded in $[\hat{T}_{\mu \rho}]_R$, we only need to determine the multiplicative renormalization $Z_{\delta}$ of the operator $\hat{\delta}_{x,\rho}$, which generates an infinitesimal translation on a generic probe $\phi_t(y)$.
In this paper, we will focus only on the numerical determination of the ratios $Z_i/Z_{\delta}$

\section{Setup}

At finite lattice spacing the renormalized energy momentum tensor is defined by the following operator mixing

\begin{gather*}
[\hat{T}_{\mu \rho}]_R=Z_1\left[\hat{T}^{[1]}_{\mu \rho}-\langle \hat{T}^{[1]}_{\mu \rho} \rangle_0\right]+Z_3\hat{T}^{[3]}_{\mu \rho}+Z_6\hat{T}^{[6]}_{\mu \rho}\quad,
\end{gather*}
where

\begin{gather}
\hat{T}^{[1]}_{\mu \rho}=-\frac{2}{g_0^2}\delta_{\mu \rho}\sum_{\sigma \tau}\tr[\hat{F}_{\sigma \tau}\hat{F}_{\sigma \tau}] \label{op_mixing1}\quad,\\
\hat{T}^{[3]}_{\mu \rho}=-\frac{2}{g_0^2}\delta_{\mu \rho}\sum_{\sigma}\tr\left[\hat{F}_{\sigma \mu}\hat{F}_{\sigma \mu}-\frac{1}{4}\sum_\tau\hat{F}_{\sigma \tau}\hat{F}_{\sigma \tau} \right]\label{op_mixing2}\quad,\\
\hat{T}^{[6]}_{\mu \rho}=-(1-\delta_{\mu \rho})\frac{2}{g_0^2}\sum_{\sigma}\tr\left[\hat{F}_{\sigma \mu}\hat{F}_{\sigma \mu} \right]\label{op_mixing3}\quad.
\end{gather}
Here $\hat{{F}}_{\sigma \mu}$ represents a lattice discretized form of the strength tensor (in our case, the clover definition has been adopted).
The index in square brackets labels operators that transform according to different irreducible representations of the hypercubic group.
Having three unknowns to be determined, we need at least three different probes: we choose to use the following operators, obtained as a flowed version of (\ref{op_mixing1}), (\ref{op_mixing2}), (\ref{op_mixing2}).

\begin{gather}
\phi^{[\alpha]}_{t,\mu}(x)=\hat{\partial}_\rho \hat{T}^{[\alpha]}_{\mu \rho}(x)|_{U_\mu=V_{t,\mu}} \quad \alpha=1,3,6\quad .
\end{gather}
We end up with a system of three equations that can be numerically solved 

\begin{gather}\label{system}
\sum_{\beta=1,3,6}\left[\sum_{\rho \sigma}\langle \hat{T}^{[\beta]}_{\mu \rho} (x)\partial_\mu \phi^{[\alpha]}_{t\rho}(x)\rangle \right]\frac{Z_{\beta}}{Z_\delta}=\sum_{\rho}\langle \hat{\delta}_{x,\rho}\phi^{[\alpha]}_{t,\rho}(x)\rangle \quad,
\end{gather}
the solution consisting in the values of $Z_{\beta}/Z_{\delta}$.

We anticipate that the setup given by (\ref{system}) is useful only at finite lattice spacing: in the continuum, the contributions from the flowed versions of $T^{[3]}$ and $T^{[6]}$ are related for $O(4)$ rotational symmetry, so the system is underdetermined.
This is consistent with the fact that, at zero lattice spacing, there are only two linearly independent, gauge invariant, local vector probes with mass dimension 5.

\section{Numerical results}

We show the preliminary results coming from the solution of system (\ref{system}).
The geometric setup consists of a lattice with open-SF boundary conditions along the time direction and periodic along the spatial ones.
The expectation values in (\ref{system}) have been measured positioning both the probe and the energy momentum tensor at $x_0=T/2$
and averaging over the spatial extensions.

At zero flow time, the TWI are affected by a contact term arising from the coalescence of probe and energy momentum tensor.
At positive flow time, we would ideally distinguish three different regions.
The first one is a small flow time window characterized by large lattice artefacts, representing the remnant of the divergence at $t=0$: these artefacts are supposed to vanish when continuum limit is approached.
After these effects have died out, the data obtained for the ratios $Z_i/Z_\delta$ should exhibit a plateau as a function of the flow time:
this is the second region of flow times, where the values of the
coefficients $Z_{i}/Z_{\delta}$ could be extracted.
Finally, we would have a big flow time window where the solution of the system would get more and more noisy due to signal depletion. 

\begin{figure}[ht]
\begin{centering}
  \includegraphics[scale=0.4]{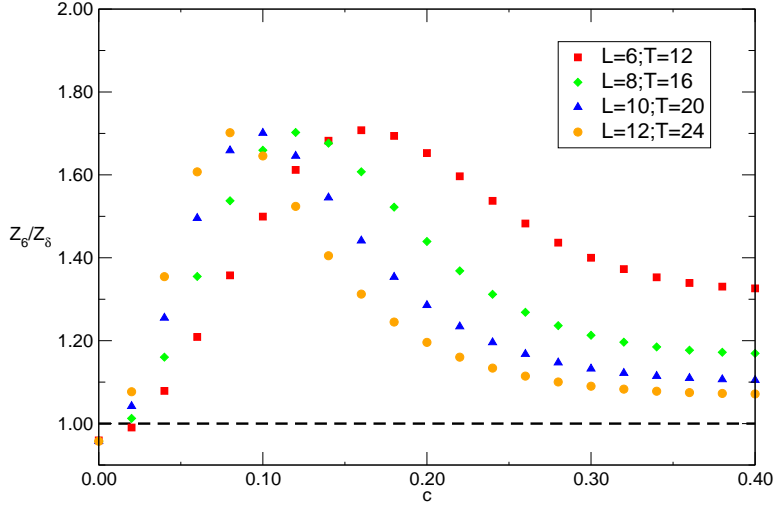}
\caption{$Z_6/Z_\delta$ tree level values  in terms of $c=\sqrt{8t/T}$.}
\label{tree}
\end{centering}
\end{figure}

The plot in figure (\ref{tree}) represents the tree level contribution to $Z_6/Z_\delta$ for values of $c$ raising from $0$ to $0.4$, $c$ being the ratio between the flow smearing radius and the lattice temporal extension.
The measurements have been taken at different, increasing volumes: if we keep the value of $c$ fixed, this is exactly equal to send $a$ to zero.
Here we see remnant effects of the divergence running from $c=0$ to
$c\simeq 0.25$, the plateau starting around $c\simeq0.3$: as continuum limit is approached, we can see how $Z_6/Z_\delta$ move towards its expected value, i.e. $Z_6/Z_\delta=1$.

\begin{figure}[ht]
\begin{centering}
 \includegraphics[scale=0.4]{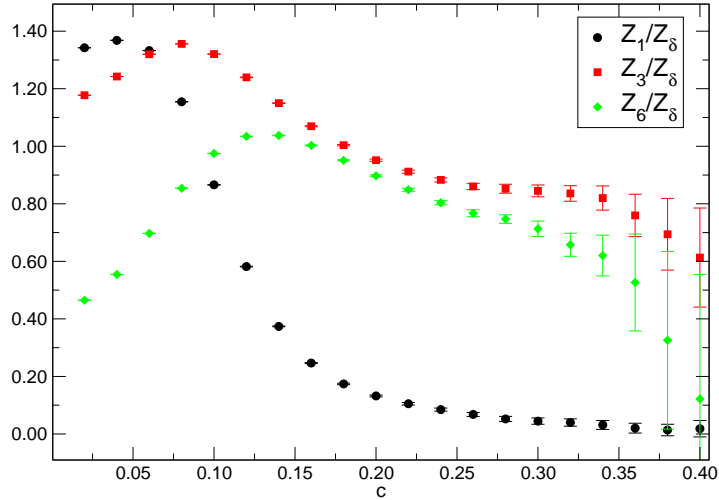}
\caption{Values of $Z_{1,3,6}/Z_{\delta}$ in terms of $c=\sqrt{8t/T}$.}
\label{constants}
\end{centering}
\end{figure}

Figure (\ref{constants}) shows the numerical values of $Z_{1,3,6}/Z_{\delta}$ from simulations on a $16\times8^3$ lattice using the same values of $c$ as the tree level case.
The measurements have been carried out using a code based on
\href{http://luscher.web.cern.ch/luscher/openQCD}{openQCD},  with
Wilson action and $\beta=6.0056$ ($a\simeq0.37 fm$)
: the statistics is about $4.4\times 10^4$ measurements.

For $Z_{3,6}/Z_\delta$, the contact term effects seem to occupy the same flow time interval as the tree level case: the main difference here consists in the loss of precision around $c=0.3$ and the impossibility to locate a real plateau of points. 

\begin{figure}[ht]
\begin{centering}
  \includegraphics[scale=0.34]{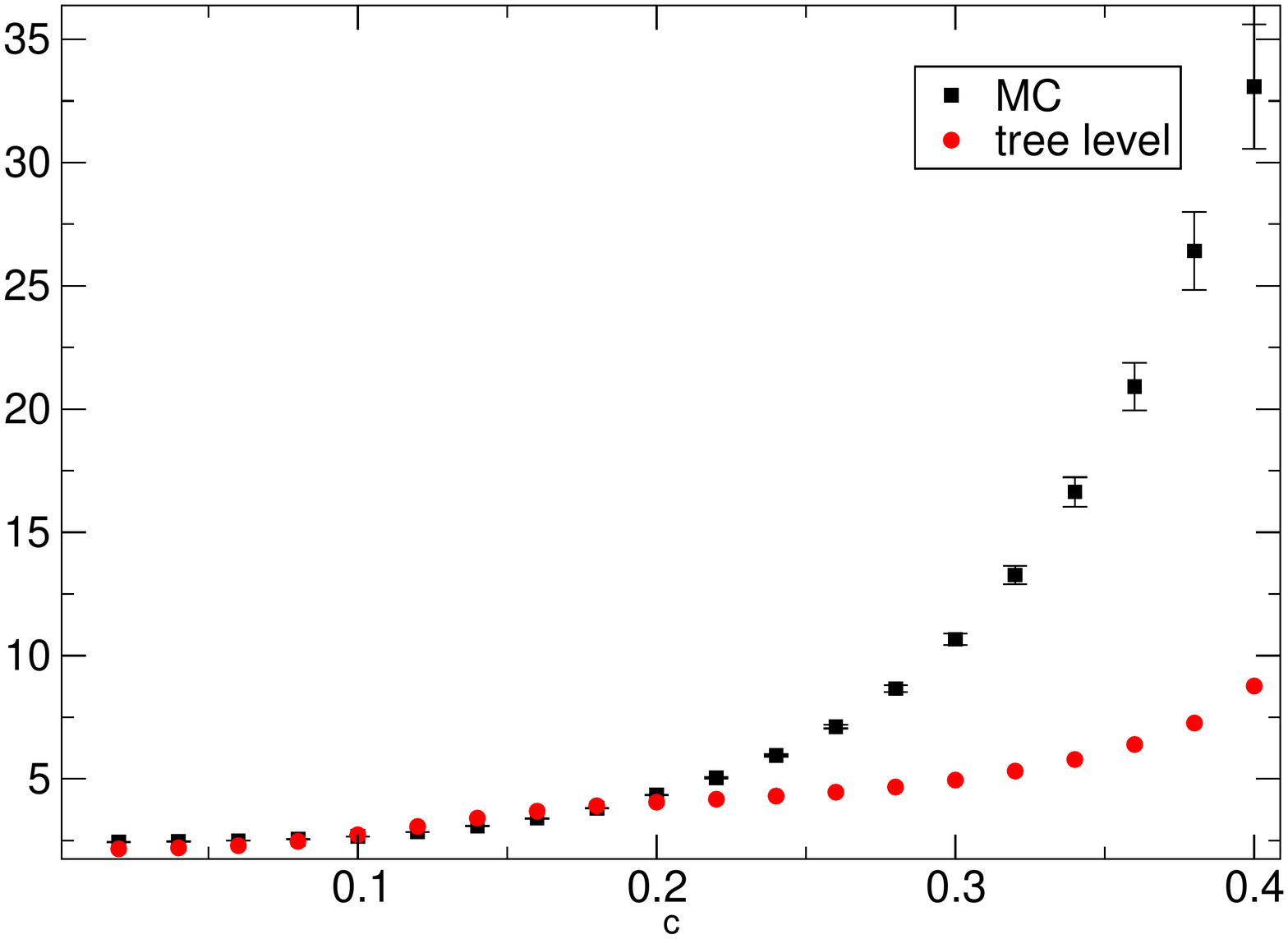}
\caption{Condition number of the matrix defined in Eq. ({\protect\ref{system}}). The red, and black,  points show the result for the tree, and the MC evaluation of the matrix respectively. }
\label{cond_numb}
\end{centering}
\end{figure}

This is consistent with the behaviour of the condition number of the matrix defined on l.h.s. of equation (\ref{system}), shown in figure (\ref{cond_numb}). As the flow time increases and signal gets worse, the condition number rapidly grows, allowing the solution to have larger fluctuations.
This in turn affects the precision with which the coefficients $Z_{1,3,6}/Z_{\delta}$ can be numerically determined.
Comparing the results coming from simulations with the ones of tree level calculations, we can see how numerical precision can make the difference in containing the values of the condition number.

Another interesting phenomenon is the behaviour of the coefficient $Z_1/Z_{\delta}$, which shows itself to be compatible with zero when $c>0.25$: however, tree level calculations showed that $Z_1/Z_{\delta}$ is $3.5\times10^{-2}$, then its determination from simulations will probably require a different, ad hoc strategy.

\section{Conclusion}

We have shown that observables at positive flow time can be used for probing Ward identities that define the renormalized energy momentum tensor.
Given a set of linearly independent probes, at finite lattice spacing, it is possible to build a system of equations whose numerical solution consists in the renormalization constants $Z_{1,3,6}/Z_{\delta}$.
However, the results of the previous section clearly tell us that the
adopted setup is not the optimal one for extracting such
coefficients in a region safe from power divergence effects, and with a good statistical accuracy.

Future work will focus on using sets of linearly independent probes with higher mass dimension.
It could be also worth to determine the values of $Z_{1,3,6}/Z_{\delta}$ through an over-constrained linear system.

\section{Acknowledgement's}
Antonio Rago is supported by the Leverhulme Trust (grant RPG-2014-118) and STFC (grant ST/L000350/1).
Luigi Del Debbio is supported by STFC, grant ST/L000458/1, and the Royal Society, Wolfson Research Merit Award, grant WM140078.
The numerical computations have been carried out using resources of the Dirac2 facilities  and of the HPCC Plymouth.

\end{document}